\documentclass[twocolumn,aps,amssymb,prb]{revtex4-1}
\usepackage[dvips,final]{graphicx}
\usepackage{color}

\usepackage{dcolumn}
\usepackage{bm}

\begin{document}

\title{On the low-field Hall coefficient of graphite}

\author{P. Esquinazi}\email{esquin@physik.uni-leipzig.de}
\author {J. Kr\"uger}
\author{J. Barzola-Quiquia}
\affiliation{Division of Superconductivity and Magnetism, Institut
f\"ur Experimentelle Physik II, Universit\"{a}t Leipzig,
Linn\'{e}stra{\ss}e 5, D-04103 Leipzig, Germany}

\author{R. Sch\"onemann, T. Herrmannsd\"orfer}
\affiliation{Hochfeld-Magnetlabor, Helmholtz-Zentrum
Dresden-Rossendorf, Germany}

\author{N. Garc\'ia}
\affiliation{Laboratorio de F\'isica de Sistemas Peque\~nos y Nanotecnolog\'ia,
Consejo Superior de Investigaciones Cient\'ificas, E-28006 Madrid, Spain}


\begin{abstract}
We have measured  the temperature and magnetic field dependence of
the  Hall coefficient ($R_{\rm H}$) in three, several micrometer
long multigraphene samples of thickness between $\sim 9~$to $\sim
30$~nm in the temperature range 0.1 to 200~K and up to 0.2~T
field. The temperature dependence of the longitudinal resistance
of two of the samples indicates the contribution from embedded
interfaces running parallel to the graphene layers. At low enough
temperatures and fields $R_{\rm H}$ is positive in all samples,
showing a crossover to negative values at high enough fields
and/or temperatures in samples with interfaces contribution. The
overall results are compatible with the reported superconducting
behavior of embedded interfaces  in the graphite structure
 and
 indicate that the negative low magnetic field Hall coefficient is not
intrinsic of the ideal graphite structure.
\end{abstract}

\maketitle

\section{Introduction}
\label{int}
The Hall effect is a  fundamental transport property of metals and
semiconductors. It can provide information on the carrier
 densities as well as on other interesting features of the electronic band
structure. Surprisingly and in spite of considerable work in the
past, the Hall coefficient ($R_{\rm H}$) of graphite, a highly
anisotropic material composed by a
 stack of graphene layers with Bernal stacking order
(ABA...), in particular the temperature and  magnetic field
dependence of $R_{\rm H}$ reported in literature is far from
clear. For example, early data on the low-field Hall coefficient
obtained in single-crystalline natural graphite (SCNG) samples
showed that it is {\em positive}  at fields below  and negative
above $\sim 0.5~$T  at a temperature $T = 77~$K \cite{sou58},
suggesting that holes are the majority carriers. This result
appears to be at odd to several other studies on the graphite band
structure obtained in highly oriented pyrolytic graphite (HOPG)
samples
\cite{kelly,gru08,zhou061,zhou062,lee08,sug07,sch09,orl08JP,gon12}
that suggest that electrons are the majority carriers, unless one
argues in terms of different mobilities of the majority carriers,
an interpretation that was used indeed in the past.

The difference between the reported Hall coefficient obtained in
the SCNG and different HOPG samples  was the subject of a short
paper in 1970 where the authors concluded that the positive
low-field Hall coefficient is observed for samples with long
enough mean free path, i.e. less lattice defects, whereas the
negative sign results from boundary scattering in HOPG samples due
to the relatively small single crystalline grains \cite{coo70}. At
high enough applied magnetic fields or high enough temperatures,
this coefficient, however, turned  negative \cite{sou58,coo70}. A
positive low-field Hall coefficient was already reported in 1953
for graphite samples with small crystal size (less annealing
temperature); it decreased with temperature and changed sign at a
sample dependent temperature \cite{kin53}. When the crystalline
grain size  in the samples was larger than $\sim 0.5~\mu$m, the
Hall coefficient was always negative, at least at $T \ge 77~$K
\cite{kin53}. Similar results were obtained in carbons and
polycrystalline graphite samples with different crystal size in
Ref.~\onlinecite{mro56}, where the authors recognized further that
the Hall coefficient was highly dependent on the {\em alignment}
of crystallites in the samples. Note that  these two last reports
\cite{kin53,mro56} are in apparent contradiction to the
relationship between crystal size and positive sign of $R_{\rm H}$
given in Ref.~\onlinecite{coo70}. It is therefore suggestive that
one extra parameter related to the alignment of the crystalline
grains in the samples could provide a hint to solve this
contradiction.

In the studies of Ref~\onlinecite{bra74} a positive Hall
coefficient was reported at 4.2~K that became negative at $\mu_0H
\ge 0.05$~T for different graphite samples. In that work
\cite{bra74} the positive low-field Hall coefficient was explained
within the two-band model arguing that it is due to the higher
mobility of the majority holes in comparison with the mobility of
the majority electrons. However, to understand its behavior as a
function of field and temperature, three types of carriers had to
be introduced in the calculations \cite{bra74}. The low-field
coefficient of different Kish graphite samples was reported in
Ref.~\onlinecite{osh82}. For the ``best" Kish sample, defined as
the one with the largest resistivity ratio $\rho(300)/\rho(4.2)$,
the Hall coefficient was positive at low fields and turned to
negative at $\mu_0 H \simeq 0.6~$T at 4.2~K, similarly to the
results for some of the graphite samples reported in
Ref.~\onlinecite{bra74}. The temperature dependence of the
zero-field Hall coefficient for the ``best" Kish specimen was
interpreted \cite{osh82} taking into account the trigonally warped
Fermi surfaces in the standard Slonczewski-Weiss-McClure's model
\cite{sw58,mcc2}. Interestingly, the lesser the perfection of  the
Kish graphite samples the larger was the field where the Hall
coefficient changed sign \cite{osh82}.

Recently published Hall measurements in micrometer small and thin
graphite flakes, peeled off from HOPG samples,  showed a positive and nearly field
independent  Hall coefficient at  $T = 0.1~$K up to 8~T
applied fields \cite{bun05}. A positive Hall coefficient
was also observed in similar graphite flakes at $77 \le T \le
300~$K, which decreased with temperature, it was field independent to  $\mu_0H \simeq 1$~T, decreasing at higher fields
\cite{van11}. Interestingly, both results \cite{bun05,van11} are
in rather good quantitative agreement with the result for the
low-field Hall coefficient reported 56 years ago for the bulk SCNG
\cite{sou58}, in clear contrast to reports in HOPG bulk
samples \cite{kelly,coo70,yakovprl03,kempa06,kopepl06,sch09}.

Further studies  on bulk HOPG samples showed the existence of an
anomalous Hall effect and a negative Hall coefficient at low
fields, interpreted as the result of a magnetic field induced
magnetic excitonic state \cite{kopepl06}. Also the quantum Hall
effect (QHE) (with electrons as majority carriers) has been
reported in some bulk HOPG samples at high enough fields
\cite{yakovprl03,kempa06}. But the QHE in graphite as well as
other interesting features of the Hall effect behavior like the
hole-like contribution with zero mass \cite{yakadv07} in bulk HOPG
samples, appear to be strongly sample dependent
\cite{sch09}. A short resume of the literature results can be
seen in Table~\ref{Tab.1}. Evidently, all these, apparently contradictory
results indicate us that we need a re-evaluation of the sign,
temperature and field behavior of the Hall coefficient. The whole
reported studies show us that we do not know with certainty, which is
the {\em intrinsic} value of the Hall coefficient in ideal
graphite and which is the origin of all the observed
differences between samples of different origins and microstructure.

\begin{widetext}

\begin{table}
\squeezetable
\label{Tab.1}
\begin{ruledtabular}
\begin{tabular}{l|c|c|c|c|c|l}
$\bf{Ref.}$&$\bf{year}$&$\bf{sample}$&$\bf{T-range}$&$\bf{field-range}$&$\bf{R_{\rm H}}$&$\bf{comment}$\\
\colrule
\colrule
\onlinecite{kin53} & 1953 & PCG & $\ge 4.2~$K & $< 1~$T & positive & for small grains only$^3$\\
& & & & & &\\
\onlinecite{mro56} & 1956 & PCG & 77~K/300~K & 0.65~T & positive & for small grains only$^4$\\
& & & & & &\\
\onlinecite{sou58} & 1958 & SCNG & $\le 77~$K & $<$ 0.5~T  & positive & negative at large fields and temperatures\\
& & & & & &\\
\onlinecite{coo70} & 1970 & HOPG/SCG &77~K & $< 0.1~$T & positive & negative upon mean free path$^2$\\
& & & & & &\\
\onlinecite{bra74} & 1974 & $(\dag)$ & 4.2 K & $<$  50 mT & positive & negative at higher fields\\
& & & & & &\\
\onlinecite{osh82} & 1982 & Kish graphite & 4.2 K & $<$ 600 mT &
positive & ${R_{\rm H}(H)}$ changes sign$^5$\\
& & & & & &\\
\onlinecite{yakovprl03} & 2003 & HOPG &0.1~K$ \le T \le 20~$K &$ \le 9~$T & negative & quantum Hall effect (QHE)$^6$ \\
& & & & & &\\
\onlinecite{bun05} & 2005 & graphite flakes$^1$ & 0.1 K & $<$ 8 T & positive &\\
& & & & & &\\
\onlinecite{kempa06} & 2006 & HOPG &2~K, 5~K & $\le 9~$T & negative & quantum Hall effect (QHE)$^6$\\
& & & & & &\\
\onlinecite{kopepl06} & 2006 & HOPG &  0.1~K$\le T \le 100~$K & $\le 0.5~$T& negative & Anomalous Hall effect\\
& & & & & &\\
\onlinecite{sch09} & 2010 & NG/HOPG & 10~mK & $<$ 10~T & negative & ($\ast$) \\
& & & & & &\\
\onlinecite{van11} & 2011 & graphite flakes$^1$ & $77 \le T \le 300$~K & $<$ 1 T & positive &\\
& & & & & &\\
\end{tabular}
\end{ruledtabular}
\caption{Selected publications that report on the  Hall
coefficient $R_{\rm H}$ of different graphite samples at different
fields and temperatures. NG: natural graphite, SCNG: single
crystal natural graphite, PCG: polycrystalline graphite.
$^1$Peeled off from HOPG. $^2{R_{\rm H}}>0$ for long mean free
path, ${R_{\rm H}} < 0 $ for short mean free path or at higher
fields. $^3$Negative for grain size $ > 0.5 \mu$m. Crossover from
positive to negative at a sample dependent temperature.
$^4$Positive $R^0_{\rm H}$ for small grains, negative for grain
size $> 0.5 \mu$m, and strong orientation dependence of $R^0_{\rm
H}$. $^5$ Turns negative at large field for samples with imperfect
structure.$^6$With electrons as majority carriers. ($\ast$): No
Hall data shown at low fields. ($\dag$): The samples were obtained
by crystallization from a solution of carbon in iron and then
purified at 2000$^{\circ}$C in a flux of chlorine. The platelets
had a thickness of $\sim 100~\mu$m~$ \lesssim d \lesssim 9~$mm. }
\end{table}

\end{widetext}

In this work, we argue that  one main reason for the observed
differences of the Hall coefficient between samples is related to
the existence of two dimensional (2D) interfaces
\cite{ina00,bar08,gar12}. Moreover, in some of them Josephson
coupled superconducting regions exist, oriented parallel to the
graphene layers of the graphite matrix
\cite{bal13,schcar,esqarx14}.  The interfaces in graphite, whose
contribution to the Hall effect we discuss in this work, are grain
boundaries between crystalline domains with slightly different
orientations. Those crystalline domains and the two-dimensional
borders between them, can be recognized by transmission electron
microscopy when the electron beam is applied parallel to the
graphene planes of graphite, see e.g. Fig.~1 in
Ref.~\onlinecite{esqarx14}, Fig.~1 in Ref.~\onlinecite{bar08} or
Figs.~2.2 and 2.9 in Ref.~\onlinecite{ina00}.  The interfaces can
be located at the borders of slightly twisted crystalline Bernal
stacking order regions (ABA...) or between regions with Bernal and
rhombohedral staking order (ABCAB...) regions. They can be
recognized usually by a certain gray colour in the TEM pictures
\cite{esqarx14,bar08}. From TEM pictures we obtain that the
distance between those interfaces can be between $\sim 30~$nm  and
several hundreds of nm upon sample \cite{bar08,schcar}. Therefore,
the thinner the graphite sample the lower the probability to have
interfaces and to measure their contribution to any transport
property.

The twist angle $\theta_{\rm twist}$, i.e., a rotation with
respect to the $c-$axis between single crystalline domains of
Bernal graphite, may vary from $\sim 1^\circ$ to $< 60^\circ$
\cite{war09} while the tilting angle of the grains with respect to
the $c$-axis $\theta_{\rm c} \lesssim 0.4^\circ$ for the best
oriented pyrolytic graphite samples. As emphasized in
Ref.~\onlinecite{esqarx14}, in case the twist angle is small enough,
the grain boundary can be represented by a system of screw
dislocations or a system of edge dislocations if the misfit is in
the c-direction with an angle $\theta_{\rm c}\neq 0$. A system of
dislocations at the two-dimensional interfaces or topological line
defects can have a large influence in the dispersion relation of
the carriers \cite{fen12,San-Jose2013} and trigger localized
high-temperature superconductivity \cite{vol14}.

There have been several theoretical studies predicting high
temperature superconductivity at the rhombohedral (ABC) graphite
surface \cite{kop11,kop13,vol13} or at interfaces between
rhombohedral and Bernal (ABA) graphite \cite{mun13,kop12}. We note
that rhombohedral graphite regions were also recognized embedded
in bulk HOPG samples \cite{lin12,hat13}. Theoretical studies
indicate an unusual dependence of the superconductivity at the
surface of
 rhombohedral graphite  or at the interfaces between rhombohedral
 and Bernal (ABA) graphite in multilayer graphene on doping \cite{mun13}.
 Furthermore, calculations indicate that high-temperature
surface superconductivity survives throughout the bulk due to the
proximity effect between ABC/ABA interfaces where the order
parameter is enhanced \cite{mun13}. Following experimental results
that indicate the existence of granular superconductivity at
certain interfaces in bulk HOPG samples of high grade
\cite{bal13,schcar,bal14I}, it is then appealing to take the
contribution of the interfaces in the behavior of the Hall effect
into account. \color{black} Doing this we are able to interpret
several results from literature as well as the Hall coefficient
obtained from three, micrometer small graphite flakes described
below. The characteristics of the embedded interfaces, as for
example their size or area \cite{bal14I} or the twist angle
between the two Bernal graphite blocks \cite{esqarx14} can have a
direct influence on the temperature and magnetic field dependence
of the Hall coefficient of a graphite sample with interfaces.
Furthermore, the alignment dependence reported earlier
\cite{mro56} can be understood arguing that the interfaces are
created or get larger in area the higher the alignment of the
grains. From the presented work we can conclude that the intrinsic
low magnetic field Hall coefficient of ideal graphite is positive, i.e.
hole like. It can change to negative in samples with embedded
interfaces at fields and temperatures high enough to influence
their contribution. On the other hand one expects that if the
interfaces are not superconducting in a given sample, they will
provide an electron-like contribution to the Hall resistance,
which may or may not overwhelm the intrinsic Hall signal of
graphite.

\section{Experimental details}

\begin{figure}[]
\begin{center}
\includegraphics[width=0.9\columnwidth]{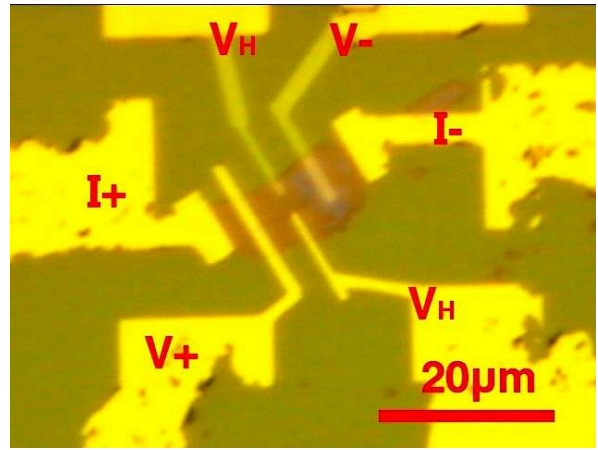}
\caption[]{Optical microscope picture of sample S1 with the two
current and four voltage gold contacts after lithography
development and evaporation of Pt/Au contacts.}\label{foto}
\end{center}
\end{figure}

The graphite flakes we have measured were obtained by a rubbing
method on Si-SiN substrates using a bulk HOPG sample of ZYA grade
from Advanced Ceramics Co. These samples  show, in general, well
defined quasi-two dimensional interfaces between Bernal graphite
structures with slightly different orientation around the
$c-$axis. Their distance in the $c-$axis direction is sample
dependent and in general $< 500$~nm and of several micrometer
length in the ($a,b$) plane \cite{ina00,bar08,schcar}.  The Pt/Au
contacts for longitudinal and transverse Hall resistance
measurements were prepared using electron beam lithography. The
samples with their substrates were  fixed on a chip carrier.
Further details on the preparation can be read in \cite{bar08}. We
have measured three samples labeled S1, S2 and S3 with similar
lateral dimensions but with thickness: $~9 \pm 1~$nm (S2), $~20
\pm 4~$nm (S1) and $~30 \pm 3~$nm (S3). An example of one of the
samples is shown in Fig.~\ref{foto}.

The transport measurements were carried out using the usual
four-contacts methods in a conventional He$^4$ cryostat and  two
of the samples (S1,S2) were also measured in a dilution
refrigerator. Static magnetic fields were provided by
superconducting solenoids applied always parallel to the $c-$axis.
The longitudinal and transverse resistances were measured using a
low-frequency AC bridge LR700 (Linear Research). After checking
the ohmic response in both voltage electrodes of the samples, all
the measurements were done with a fixed current of $1~\mu$A, which
means a dissipation $ \mathrm{d}Q/\mathrm{d}t < 0.1~$nW to avoid
self heating effects. In this work we focus on the  Hall
coefficient obtained at fields $\mu_0 H \lesssim 0.2~$T.

\section{Results and discussion}

\subsection{Temperature dependence of the longitudinal resistance}

\begin{figure}[]
\begin{center}
\includegraphics[width=1\columnwidth]{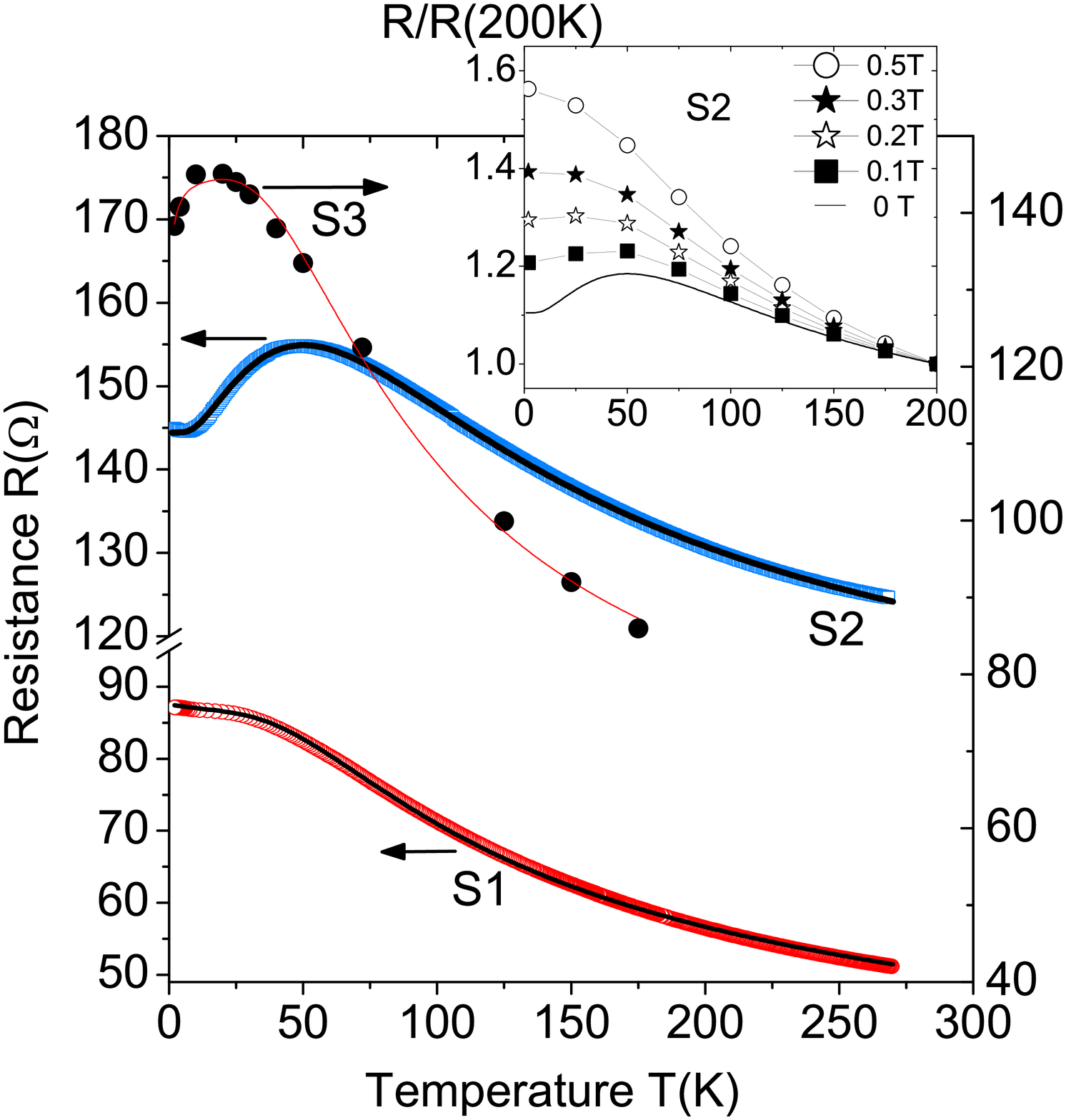}
\caption[]{Resistance vs. temperature for the three multigraphene
samples discussed in this work. The measurements were made without
any applied field. The continuous lines were calculated following
the parallel-resistor model from Ref.~\onlinecite{gar12}, see
text. The inset shows the normalized resistance vs. temperature at
different fields applied normal to the interfaces for sample S2.
}\label{fig1}
\end{center}
\end{figure}

Figure~\ref{fig1} shows the electrical longitudinal resistance of
the three samples vs. temperature at zero applied field. Following
the results and the discussion exposed in Ref.~\onlinecite{gar12},
we assume that ideal graphite is a narrow-gap semiconductor. The
observed semiconducting-like temperature dependence in the
longitudinal resistance has been also reported recently for thin
graphite flakes
 \cite{van11}. We assume that any
deviation from a semiconducting-like dependence in the
longitudinal resistance is due to extrinsic contributions. The
observed behavior in our samples is similar to that already
published \cite{bar08} and it can be understood taking into
account the contributions  of the semiconducting graphene layers
in parallel to that from the embedded interfaces  and of the
sample surfaces (open and with the substrate) \cite{gar12}. The
interfaces' contribution is responsible for the maximum in the
resistance observed at $\sim 50~$K and $\sim 20~$K for samples S2
and S3, respectively. We speculate that the sample surfaces are
responsible for the saturation of the resistance at low
temperatures, as in S1 for example. \color{black} The embedded
interface contribution appears to be weaker in sample S1 than in
the other two samples, therefore we expect for this sample a Hall
coefficient with less extrinsic contributions than for the other
two.

With a simple parallel-resistor model one can understand
quantitatively the measured temperature dependence using different
weights between the parallel contributions  following the
relation  \cite{gar12}:
\begin{equation}
R(T) = (R_i^{-1}(T) +
R_b^{-1}(T))^{-1}\,,
\label{eqrespar}
\end{equation}
see Fig.~\ref{fig1}. The bulk, intrinsic contribution of graphite
is semiconducting-like
\begin{equation}
R^{-1}_b(T) \simeq (A
\exp(E_g/2k_BT))^{-1} + R_d^{-1}\,,
\end{equation}
and the one from the interfaces
and surfaces can be simulated following
\begin{equation}
R_i(T) = R_0 + R_1 T +
R_2\exp(-E_a/k_BT)\,,
\end{equation}
whereas the parameters $A, R_j (j = 0,1,d)$ are free. The constant
term $R_d$  prevents an infinite resistance from the bulk
contribution, which is physically related to defects or (a)
surface band(s). $E_g, E_a$ denote the semiconducting gap and an
activation energy; their values depend on sample within the range
$250~$K~$ \lesssim E_g \lesssim 500~$K and $10~$K~$ \lesssim E_a
\lesssim 50~$K \cite{gar12}. The linear in temperature term ($R_1
T$) could be negative or positive and it is taken as a guess for
the contribution of the surfaces and/or metallic-like interfaces.
The thermally activated function can be interpreted as the
contribution of non-percolative, granular superconducting regions
inside the internal interfaces embedded in the graphite matrix
\cite{gar12}, similar to that observed in granular Al in a Ge
matrix \cite{sha83}, for example. Transport \cite{bal13} as well
as magnetization \cite{schcar} measurements support the existence
of granular superconductivity and  Josephson coupling between
superconducting regions at these interfaces.

At fields of the order of 0.1~T applied perpendicular to the
interfaces, the metallic-like behavior (at $T < 50~$K) starts to
vanish, see the results of sample S2 in the inset of
Fig.~\ref{fig1}, as example. This behavior  is assigned to the
field-driven superconductor- (or metal-) insulator transition
\cite{kempa00,yakovprl03,tok04,heb05}. Because this behaviour is
absent in thin enough graphite flakes \cite{bar08,gar12} or in
thicker graphite samples without well defined interfaces, the
field-driven transition is not intrinsic of ideal graphite and
should not be interpreted in terms of band models for graphite
with Bernal stacking order \cite{tok04,heb05}. If this
field-driven transition is compatible with the existence of
Josephson coupled superconducting regions at the interfaces
\cite{bal13,gar12,schcar}, we therefore expect that the Hall
coefficient should be influenced at similar applied fields, as we
show below.

\subsection{Temperature dependence of the low-field Hall coefficient}

The low-field Hall  coefficient is defined as the $R^0_{\rm H} = d
\lim_{H \rightarrow 0} r_{\rm H} /\mu_0 H$, where $r_{\rm H}$ is
the Hall resistance and $d$ the measured thickness of the sample.
Figure \ref{fig5} shows the  temperature dependence of $R^0_{\rm
H}(T)$ for the three samples measured in this work. We note that
it changes sign at $\sim 45~$K and $\sim 70~$K for samples S3 and
S2. For sample S1, $R^0_{\rm H}(T)$   remains positive in the
whole measured temperature range. The observed behavior of
$R^0_{\rm H}$ for sample S1  as well as its absolute value are
similar to the recently reported ones for a mesoscopic graphite
flake of similar thickness \cite{van11}. Due to the large
dispersion of Hall coefficients and their variation with
temperature found in literature (see Table I in Sec.~\ref{int}),
this agreement is remarkable and support early results on a
positive $R^0_{\rm H}$ at low enough fields and temperatures for
graphite \cite{sou58,coo70}.

All the three samples show positive, saturating  $R^0_H(T \lesssim
25$~K), see Fig.~\ref{fig5}. The origin of the differences between the  low-field Hall
coefficients at low temperatures is not  known with certainty.
A quantitative comparison
of the absolute values of the Hall coefficient between different
graphite samples is not straightforward  because different
 samples have different
interface contributions. The density of these interfaces as well
as the number of those that have superconducting properties depend
on the sample and it is not simply proportional to the thickness
of the sample. Also different absolute values could arise from
differences between the  properties of holes and electron
carriers, see Eq.~(\ref{hallc}), whose densities and mobilities
can be influenced by defects and impurity atoms. Therefore,
differences in the absolute values of the Hall coefficient between
the samples should be taken with some care.

\subsubsection{The intrinsic Hall coefficient}

Taking into account the $T-$dependence  of the longitudinal
resistance, the results of Ref.~\onlinecite{gar12} and the fact
that $R^0_{\rm H}(T)$ is positive,  at least a two-band model
\cite{kelly} is necessary to interpret the data. According to this
model, the Hall coefficient is given by:
\begin{equation}
R^0_H = \frac{\mu_p^2 p - \mu_n^2 n}{e(\mu_p p + \mu_n n)^2}\,,
\label{hallc}
\end{equation}

where $\mu_{p,n}$ are the mobilities  for holes with density $p$
and electrons with density $n$, respectively; $e$ is the positive defined
electronic charge. Since $R^0_H>0$, then  $\mu_p^2 p > \mu_n^2 n$.

Taking into account the expected band structure of graphite and
for practical purposes, we can assume either:  (1) both mobilities
are equal and that the carrier densities are related through $p =
n + \delta > n$ with $\delta \ll p,n$, or (2) $\mu_p =
\mu_n+\delta>\mu_n$, $p= n$ and $\delta \ll \mu_p,\mu_n$. In both
cases $p + n \simeq 2p$  and Eq.~(\ref{hallc}) can be approximated
by either:
\begin{eqnarray}
R^0_{\rm H} & \simeq & \frac{\delta}{4ep^2}~\textrm{with~} \delta = p - n \label{hallc1}\\
\textrm{or} &&\nonumber\\
R^0_{\rm H} & \simeq & \frac{\delta}{2ep\mu_p}~\textrm{with~}
\delta = \mu_p - \mu_n \,. \label{hallc2}
\end{eqnarray}

Obviously, the use of the one-band model equation, i.e. $R^0_H =
1/pe$, provides incorrect values for the carrier concentration.
For example, in Ref.~\onlinecite{van11} the authors obtained
$p(77~$K) $\simeq 7.7 \times 10^{19}~$cm$^{-3}$ or $p_{2D}(77~$K)$
\simeq 2.5 \times 10^{12}~$cm$^{-2}$ per graphene layer.  Using
the same one-band model we would obtain for sample S1, $p_{2D}
\simeq 8 \times 10^{12}~$cm$^{-2}$, a value four orders of
magnitude larger than the one obtained for similar samples using a
model independent constriction method \cite{dus11}. Moreover, for
such large $p$ values the use of Eq.~(\ref{hallc1}) would give
$\delta \gtrsim p$ using the measured $R^0_H$. However, if we take
the carrier concentration $p_{2D}(75~$K$) \simeq 5 \times
10^{8}$cm$^{-2}$ (or $p_{3D} \simeq 1.6 \times 10^{16}$cm$^{-3}$)
from Ref.~\onlinecite{dus11}, using Eq.~(\ref{hallc1}) we obtain
$\delta/p = 2.5 \times 10^{-4}$, indicating a very small
difference between electron and hole carrier densities. Or using
Eq.~(\ref{hallc2}) we obtain $\delta/\mu_p \simeq 1 \times
10^{-4}$, indicating a very small difference between the electron
and Hall mobilities.

\begin{figure}[]
\begin{center}
\includegraphics[width=1.00\columnwidth]{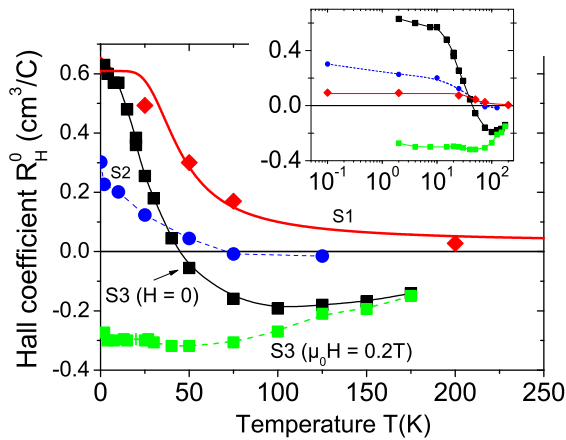}
\caption[]{Temperature dependence of the low-field Hall
coefficient for the three studied samples. For sample S3 (black
squares) we show also the Hall coefficient (green squares)
obtained at a field $\mu_0 H = 0.2~$T applied normal to the
interfaces. The data from sample S1 in the main panel of the
figure were multiplied by a constant factor of 6.7. The inset
shows the same data but in a logarithmic temperature scale. The
line through the S1 data points follows the function $R^0_{\rm H}
= (11^{-1} + (4\times 10^{-3}
 \exp(300/(2T)))^{-1})^{-1}$. All other lines
are only a guide to the eye.}\label{fig5}
\end{center}
\end{figure}

For sample S1, $R^0_{\rm H}(T)$ can be understood  as the parallel
contributions \cite{pet58} from the graphene layers with $p(T),
n(T) \propto \exp(-E_g/2k_{\rm B}T) $ and a roughly temperature
independent term (from the surfaces or some internal interfaces)
that prevents the divergence of $R^0_{\rm H}(T \rightarrow 0)$.
Using the same energy gap that fits the longitudinal resistance,
we can fit $R^0_{\rm H}(T)$ in the whole $T-$range, see
Fig.~\ref{fig5}.

\subsubsection{Interface contribution to the Hall coefficient}
\label{T-dep}
The results for $R^0_H(T)$ of sample S1 and those from
ref.~\onlinecite{van11} indicate us that the sign change above
certain temperature in samples S3 and S2 should be related with
the contribution of  interfaces. Real graphite samples with
interfaces are rather complex systems in the sense that the
distribution of input electrical currents inside the sample is not
homogeneous. Without knowing this distribution and the intrinsic
conductivities of the different contributions, quantitative models
are only under certain assumptions applicable. To estimate the
interface contribution we use the model proposed in
Ref.~\onlinecite{pet58} for a bilayer, where the Hall coefficient
of the surface (in our case the interfaces $R^i_H(T)$) and bulk
($R^b_H(T))$ contribute in parallel. The total measured Hall
coefficient is given by
\begin{equation}
R_{\rm H} = \frac{d(R^b_{\rm H} \sigma^2_bd_b + R^i_{\rm H}
\sigma^2_i d_i)}{(\sigma_bd_b + \sigma_id_i)^2}\,, \label{eqnhall}
\end{equation}
where $\sigma_{b,i}$ are the conductivities of the bulk and
interface  contributions and $d_{b,i}$ the respective effective
thicknesses, i.e. the total thickness of the sample is $d = d_b +
d_i$. Taking into account the interface density in our HOPG
samples  obtained from transmission electron microscopy
\cite{bar08}, we estimate  $d_i/d_b \lesssim 10^{-2}$. In this
case Eq.~(\ref{eqnhall}) can be written as:
\begin{eqnarray}
R_{\rm H} &\sim &\frac{R^b_H + R^i_H r_2}{r_1}\,,\\
r_1 &=& (1 + r'_1)^2, r'_1 = \frac{\sigma_id_i}{\sigma_bd_b}\,,\\
r_2 &=& \frac{\sigma_i^2d_i}{\sigma_b^2d_b} = r'^2_1
\frac{d_b}{d_i}\,. \label{eqnhall2}
\end{eqnarray}
The effective parallel contribution of the interfaces can be estimated from
\begin{equation}
R^i_H \sim \frac{R_{\rm H} r_1 - R^b_H}{r_2}\,. \label{eqhallrs}
\end{equation}

In clear contrast to sample S1, the $R^0_{\rm H}(T)$ of sample S3
turns to negative at $T > 45~$K. We propose that the origin for
the sign change of $R^0_{\rm H}(T)$ increasing $T$ is due to the
extra contribution of the interfaces when the superconducting
properties of the interfaces vanish.
 At low enough temperatures the interfaces with the
 superconducting regions \cite{bal13} do not
contribute substantially to the total low-field Hall effect.
Because we are dealing here with the zero- or low-field Hall
coefficient, vortices or fluxons (and their movement) are not
expected to influence the Hall signal.

\begin{figure}[]
\begin{center}
\includegraphics[width=0.9\columnwidth]{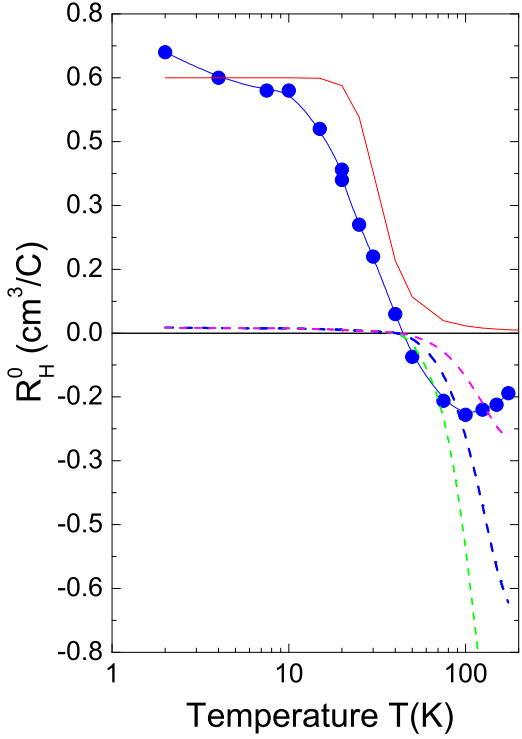}
\caption[]{Temperature dependence of the low-field Hall
coefficient for the S3 sample. The red line follows the equation
$R^b_{\rm H} = 0.03(20^{-1} + (0.1 \exp(350/(2T)))^{-1})^{-1}$
used as the bulk contribution to the Hall coefficient for that
sample. The dashed lines represent the contribution from the
interfaces $R^i_{\rm H}(T)$ calculated from Eq.~(\ref{eqhallrs})
assuming different geometrical ratios $w_b\ell_i/w_i\ell_b = 200,
300, 500$, from bottom to top lines, respectively.}\label{ri}
\end{center}
\end{figure}

For sample S3 and using  Eq.~(\ref{eqhallrs}) we assume that
$R^i_{\rm H}(T \ll 50~$K$) \simeq 0$ implying $R^0_{\rm H} r_1
\ggg R^b_{\rm H}$. If we assume further that the total low-field
Hall coefficient for this sample and at low temperatures is mainly
given by the bulk contribution, we can roughly estimate the
expected contribution from the interfaces. Using the functions
that fit the temperature dependence of the measured resistance,
see Fig.~\ref{fig1}, the related conductivities can be estimated
from $\sigma^{-1}_i \sim g_i (140 - 0.045 T + 10 \exp(-4/T))$ and
similarly for the bulk contribution
 $\sigma^{-1}_b \sim g_b (80 \exp(350/2T))$,
 where the constant prefactors $g_{b,i} \sim w_{b,i} \times d_{b,i}/ \ell_{b,i}$
 (width $\times$ thickness $/$ length)
 are due to the different effective geometry of the two contributions.
 Note that neither the samples have perfectly rectangular shape nor the internal interfaces
 are expected to follow  perfectly the measured external sample geometry.
We estimate the Hall coefficient due to the bulk contribution as
$R^b_{\rm H}(T) \sim ( 20^{-1} + (0.1 \exp(350/2T))^{-1})^{-1}$
assuming that the saturation of $R^0_{\rm H}$ we measured for this
sample, see Fig.~\ref{ri}, can be included in $R^b_{\rm H}$.
Assuming $d_b/d_i = 50$, we estimate the interface contribution to
the Hall coefficient using Eq.~(\ref{eqhallrs}) for three values
of the factor $w_b\ell_i/w_i\ell_b$ that enters in $r'_1$. The
three curves can be seen in Fig.~\ref{ri}. The uncertainty
 in the geometrical factors and in the bulk Hall contribution
 do not allow a better quantitative estimate of the interface
 contribution. Nevertheless, qualitatively the obtained results
 for $R^i_{\rm H}(T)$ appear reasonable.

\subsection{Magnetic field dependence of the Hall coefficient}
\label{B-dep}

\begin{figure}[]
\begin{center}
\includegraphics[width=1\columnwidth]{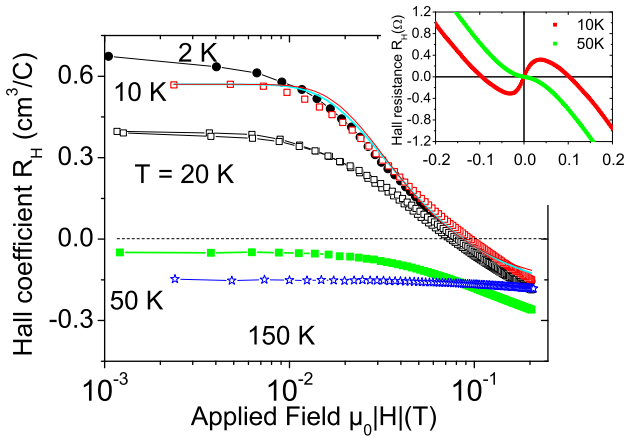}
\caption[]{Hall coefficient as a function of the absolute value of
the applied field at several temperatures for sample S3. The inset
shows the field  dependence of the Hall resistance at two
temperatures.}\label{fig4}
\end{center}
\end{figure}

We emphasize that for graphite samples with no measurable evidence
for the interface contribution, in the electrical resistivity for
example, the Hall coefficient does not depend on the field, at
least up to 1~T applied normal to the graphene planes
\cite{bun05,van11}.  Therefore, a direct way to test our
assumption that the Hall coefficient and its negative value is not
intrinsic -- but due to the extra contribution from the embedded
interfaces with superconducting regions -- can be independently
done measuring it at finite magnetic fields applied normal to the
interfaces. In this case we expect that a magnetic field will have
the same influence on the interface contribution as the
temperature. In other words, a large enough magnetic field will
destroy the coupling between the superconducting regions, or the
superconductivity itself at the interfaces and an extra,
electron-like contribution should be measurable, in principle in
the whole temperature range. Note that mostly electron-like
carriers are expected to be at the interfaces with a density of
the order of $\lesssim 10^{12}~$cm$^{-2}$. This is inferred from
Shubnikov-de Haas (SdH) oscillations in the magnetoresistance
obtained in samples with and without (or with less number of)
interfaces, see, e.g., Ref.~\onlinecite{bar08} (compare there
samples L5 and L7)  or Ref.~\onlinecite{oha01} where a clear
decrease in the amplitude of the SdH oscillations decreasing the
thickness of the samples has been reported earlier. The reason why
mostly electron-like carriers appear to be at the interfaces is
related to the nature of the interfaces themselves, a subject that
is being discussed nowadays, see Ref.~\onlinecite{esqarx14} and
Refs. therein. For example, in addition to the twist angle between
the graphite Bernal blocks forming an interface, one has extra
doping through hydrogen or carbon vacancies that influence the
carrier density and the superconducting regions at the interfaces.
For interfaces without superconducting regions we expect an
electron-like contribution to the Hall coefficient with a  weaker
field and temperature dependence.

How large should be the magnetic field to affect the coupling
between the superconducting regions or the superconductivity of
the regions itself? An estimate of this field can be directly
obtained from the longitudinal resistance and the metal-insulator
transition (MIT) observed in several high grade graphite samples,
as, e.g., in sample S2, see inset in Fig.~\ref{fig1}, or several
other samples reported in literature
\cite{yakovadv03,yakovprl03,tok04,heb05}. From all the results for
the longitudinal resistance we expect that a field of the order of
0.1~T should be sufficient to influence substantially the Hall
coefficient contribution of the interfaces. Note that the MIT of
graphite is absent for samples without or with negligible amount
of interfaces \cite{bar08,gar12,van11,bun05}. Figure~\ref{fig5}
shows the Hall coefficient of sample S3 at a field of 0.2~T. It
remains nearly temperature independent below 100~K and matches the
results obtained at low-fields at high enough temperatures.

Figure~\ref{fig4} shows the field dependence of $R_{\rm H}$ at
various constant temperatures where we recognize a transition from
positive to negative values at fields similar to those necessary
to trigger the metal-insulator transition observed in the
longitudinal resistance. The results shown in this figure clearly
indicate that a magnetic field has the same influence on the Hall
coefficient as temperature. Note also that a field of the order of
0.1~T is enough to change the sign of the Hall coefficient in the
sample with clear contribution of the interfaces.

Apart from the interface effects, one may expect a decrease of the
(positive) Hall coefficient with field, at large enough fields
when the cyclotron energy $\hbar \omega_c$ is of the order of the
energy gap $E_g$ ($\omega_c = e\mu_0H/m^\star$ and $m^\star$ the
effective electron mass, according to Ref.~\onlinecite{coh61}.
Because in graphite $E_g \lesssim 50~$meV this effect may start be
observable at $\mu_0 H > 1~$T. On the other hand, data obtained
from very thin graphite samples clearly show that the Hall
coefficient is field independent up to 10~T at $T = 0.1~$K
\cite{bun05}, or up to 1~T at $T \geqslant 77~$K \cite{van11}.
Therefore, we can clearly argue that the observed field dependence
in our sample, see Fig.~\ref{fig4}, is not intrinsic of the
graphite structure but has an extrinsic origin, similar to that
reported earlier \cite{sou58,coo70}.\\
\begin{figure}[]
\begin{center}
\includegraphics[width=1\columnwidth]{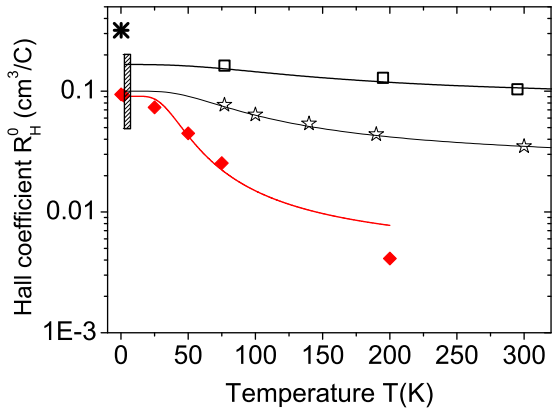}
\caption[]{Low-field Hall coefficient as a function of temperature
for sample S1 from this work (red lozenge); the line through the
points follows the equation $(11 + (0.004
\exp(300/2T))^{-1})^{-1}$. The other points are taken from
different graphite samples reported in literature. $(\star)$:
Taken from Ref.~\onlinecite{van11} for a graphite thin flake; the
line through the points follows the equation $(10 + (0.03
\exp(350/2T))^{-1})^{-1}$. $(\square)$: Taken from
Ref.~\onlinecite{kin53}; the line through the points follows the
equation $(6 + (0.15 \exp(400/2T))^{-1})^{-1}$. The vertical
dashed region at low temperatures with $0.05 \le R^0_H \le
0.2$~cm$^3$/C is from Ref.~\onlinecite{bra74}. $(\ast)$: Hall
coefficient at 100~mK obtained for a 5~nm thick graphite sample
from Ref.~\onlinecite{bun05}. }\label{fig6}
\end{center}
\end{figure}

\section{Comparison with literature and conclusion}

\begin{figure}[]
\begin{center}
\includegraphics[width=1\columnwidth]{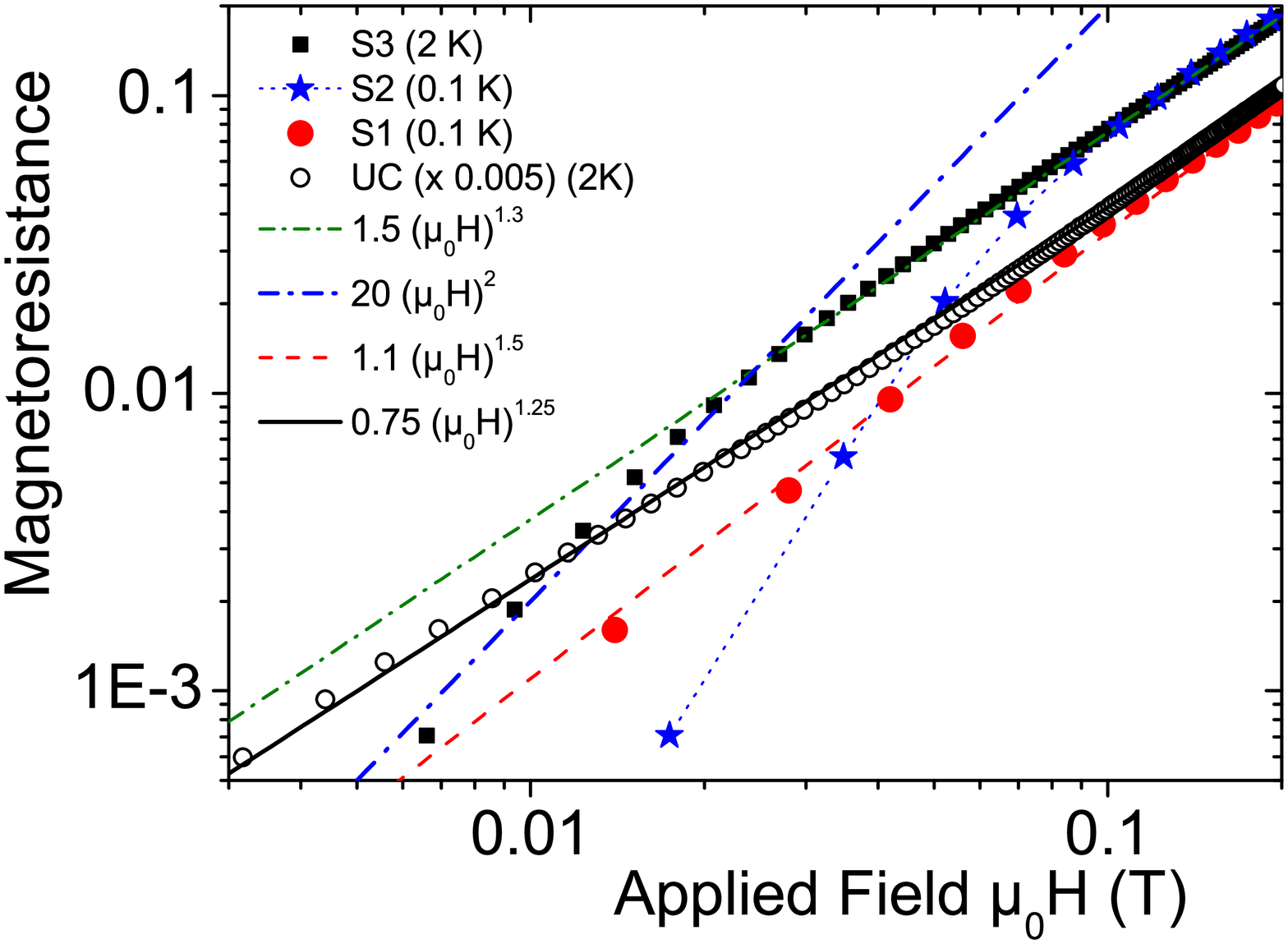}
\caption[]{Low-field magnetoresistance defined as $R(B) - R(0) /
R(0)$ vs. applied field in a double logarithmic scale, for the
three samples studied in this work. The data of samples S1 and S2
were taken at 0.1~K and of sample S3 at 2~K. In the same panel we
show the data for a bulk HOPG sample of grade A ($\circ$) obtained
at 2~K. }\label{MR}
\end{center}
\end{figure}

In a recent theoretical work\cite{pal13}, the magnetoresistance
and Hall resistivity for graphite has been calculated using the
usual 3D band structure described by the Slonczewski-Weiss-McClure
model and taking into account only some of its six free parameters
\cite{brandt88}. The obtained results indicate that at relatively
weak applied magnetic fields $< 1~$T, the magnetoresistance
increases linearly with field due to the presence of extremely
light, Dirac-like carriers. Interestingly, in the same field range
the authors found that the Hall coefficient should be positive and
proportional to $\ln|B|$. We note that a linear field
magnetoresistance was indeed reported in this field range and at
low $T$ in a large amount of graphite samples,  especially for
relatively thick graphite samples, see for example the
magnetoresistance curves for sample L7 (75~nm thick) in
Ref.~\onlinecite{bar08}. Due to the observed positive Hall effect
at low fields $ < 0.2~$T and at low temperatures, it is of
interest to check whether such a linear field dependence is
observed in the samples described in this work. Figure~\ref{MR}
shows the magnetoresistance vs. applied field below 0.2~T for the
three samples and at low temperatures. Interestingly, none of the
samples show a clear linear field dependence. The sample S1, which
has the smallest contribution from interfaces (see
Fig.~\ref{fig1}) follows approximately a $H^{1.5}$ dependence. The
other two samples, S2 and S3 tends to follow a quadratic
dependence at low enough fields changing to a $\sim H^{1.3}$
dependence at higher fields, in agreement with similar
measurements but in bulk HOPG samples \cite{kop06}. Note that the
absolute value of the magnetoresistance at a given field for our
mesoscopic samples is much smaller than in bulk samples. As an
example, we show in Fig.~\ref{MR} the results for a bulk HOPG
sample of grade A. In the depicted field region the
magnetoresistance follows $\sim H^{1.25}$ but it is 200 times
larger than in the other mesoscopic samples. This difference might
be related to the size dependence of the magnetoresistance when
the mean free path is of the order of the sample size
\cite{gon07,dus11}. The field dependence in this low field region,
however, does not seem to be affected. \color{black}

We note that in the same field range we did not see an
increasing Hall coefficient with field in any of the samples
studied here. Also, for the graphite samples reported in
Ref.~\onlinecite{van11} at $T \ge 77~$K the Hall coefficient is
constant below 1~T and decreases above. In the case of the $\simeq
1 \times 1~\mu$m$^2$ sample reported in Ref.~\onlinecite{bun05} we
note that the magnetoresistance is negligible to 10~T applied
field and $T = 0.1~$K, a result related to the ballistic behavior
of the carriers  due to their large mean free path \cite{dus11},
and the Hall coefficient does not depend on field.

A comparison of the Hall coefficient and in general of the Hall
data from literature is not straightforward because the sample
quality as well as the  existence of interfaces (of any kind) was
not provided in any of  the publications and remains, in general,
unknown. Nevertheless, we can speculate the following trends and
provide a possible answer to the works listed in Tab.~\ref{Tab.1}.
In Refs.~\onlinecite{kin53,mro56}, a positive Hall effect was
observed for small grains only. Upon preparation conditions,
smaller grains should have less
 interfaces, see section~\ref{int},  and therefore less contribution from them.
 Note that the reported dependence of $R_H$ on the {\em alignment} of the crystallites can be
 directly related to the larger probability to have  well-defined and larger
 interfaces the larger the alignment of the crystallites is. Note that effective critical temperature
 depends on the size or area of the interfaces according to recently published experimental results\cite{bal14I}.
The crossover to a negative Hall coefficient at large enough
fields and temperatures observed in
Refs.~\onlinecite{sou58,bra74,osh82,sch09} can be understood in a
similar way as shown in sections~\ref{T-dep} and \ref{B-dep}. In
Ref.~\onlinecite{coo70} the Hall coefficient was reported  to be
positive for samples with long mean free path of the carriers and
negative for samples with smaller mean free path or high fields.
The crossover to a negative $R_H$ can be understood in the same
way if some of the interfaces get normal conducting with field.
Now, the reported mean free path dependence of $R_H$ cannot be
simply interpreted in terms of interfaces contribution without
knowing the internal structure of the samples and whether there is
or not a crossover to negative $R_H$ at higher fields and
temperatures. If we assume that the carriers in the graphene
layers of graphite have much larger mobility\cite{dus11} than
those carriers at the interfaces in the normal state, we may
speculate that samples with smaller mean free path have larger
density of interfaces and therefore a negative $R_H$ should be
measured.

The negative QHE measured in the HOPG samples in
Refs.~\onlinecite{yakovprl03,kempa06} at high fields should come
from normal conducting interfaces with a relatively high density
of carriers. Note that those HOPG samples are the ones that show a
high density and well-defined two-dimensional interfaces
\cite{bar08}. That the QHE is not observed in all HOPG samples,
even when they are from the same grade\cite{sch09} (or even batch)
is related to a non-homogeneous distribution of the interfaces.
The observation of the AHE in Ref.~\onlinecite{kopepl06} is
related to defects that trigger magnetic order \cite{dim13}. Note
that even in a mesoscopic graphite sample one can find different
contributions to the transport depending how homogeneous the
sample is, see e.g., Ref.~\onlinecite{barjsnm10}. Therefore, in
this kind of samples it is difficult to measure the intrinsic
contribution coming from the ideal graphene layers. Finally, the
positive $R_H$ measured in Refs.~\onlinecite{bun05,van11} can be
expected since those samples were very probably free from
interfaces due to their small thickness.

\color{black} In general we can state that at low enough fields
and in thin enough samples with low density of interfaces, the
Hall coefficient should be closer the one from the ideal bulk
graphite than at higher fields. Therefore we show in
Fig.~\ref{fig6} low-field coefficients obtained from old and
recent publications at different temperatures, in case these data
were available. From all these data, we conclude that the
intrinsic, low magnetic field Hall coefficient of graphite appear
to be positive with a low-temperature value around $0.1~$cm$^3$/C
and a temperature dependence that follows closely that of a
semiconductor with an energy gap of the order of 400~K, in
agreement with the fits of the longitudinal resistance of
different samples \cite{gar12}.  Note that the results shown in
Fig.~\ref{fig6} were obtained from graphite samples with very
different shapes, i.e. bulk samples in
Refs.~\onlinecite{kin53,bra74} and mesoscopic samples with
different areas in Refs.~\onlinecite{bun05,van11}. From this
comparison we would conclude that the low temperature value of the
Hall coefficient does not seem to strongly  depend on the defined
Hall geometry.

\color{black}

In conclusion, taking into account the contribution of
superconducting regions at certain interfaces found in real
graphite samples, we provide a possible explanation for the
anomalous temperature and low magnetic field behavior  of the Hall
coefficient as well as for its differences  between samples of
different origins reported in the last 60 years.

We thank Yakov Kopelevich for fruitful discussion.   Part of this
work has been supported by EuroMagNET II under the EC contract
228043.



%

\end{document}